\documentclass[twocolumn]{aastex631}

\usepackage{xspace, color}
\usepackage{stmaryrd, booktabs}
\usepackage{comment}

\newcommand{\kms}{\hbox{\ensuremath{\rm{km}~\rm s^{-1}}}\xspace}
\newcommand{\Msun}{\hbox{\ensuremath{\rm M_\odot}}\xspace}

\newcommand{\AAA}{\hbox{\ensuremath{\rm \AA}}\xspace}

\newcommand{\Mtot}{\hbox{\ensuremath{M_{\rm tot}}}\xspace}

\newcommand{\sn}{SN\xspace}
\newcommand{\sneia}{SNe Ia\xspace}
\newcommand{\snia}{SN Ia\xspace}

\newcommand{\Ha}{H$\alpha$\xspace}
\newcommand{\fCo}[1]{[\ion{Co}{#1}]\xspace}
\newcommand{\fsNi}{$^{56}$Ni\xspace}

\newcommand{\Mstar}{\hbox{\ensuremath{\rm M_\star}}\xspace}
\newcommand{\logMstar}{\hbox{\ensuremath{\log_{10}\rm M_\star}}\xspace}
\newcommand{\logSFR}{\hbox{\ensuremath{\log_{10}\rm{SFR}}}\xspace}
\newcommand{\logsSFR}{\hbox{\ensuremath{\log_{10}\rm{sSFR}}}\xspace}

\newcommand{\vsep}{\hbox{\ensuremath{v_{\rm sep}}}\xspace}
\newcommand{\sne}{SNe Ia\xspace}
\newcommand{\Mv}{\hbox{\ensuremath{M_V}}\xspace}

\newcommand{\sbv}{\hbox{\ensuremath{s_{\rm BV}}}\xspace}
\newcommand{\Mni}{\hbox{\ensuremath{M_{\rm 56Ni}}}\xspace}

\newcommand{\vorb}{\hbox{\ensuremath{v_{\rm orb}}}\xspace}
\newcommand{\fFe}[1]{[\ion{Fe}{#1}]\xspace}

\newcommand{\Rwd}{\hbox{\ensuremath{R_{\rm WD}}}\xspace}

\submitjournal{MNRAS}

\begin{document}

\title{Merging White Dwarf Binaries Produce Type Ia Supernovae in Elliptical Galaxies}

\shorttitle{Bimodal SNe Ia in Massive Ellipticals}
\shortauthors{M. A. Tucker}

\author[0000-0002-2471-8442]{Michael A. Tucker}
\altaffiliation{CCAPP Fellow}
\affiliation{Center for Cosmology and AstroParticle Physics, 191 W Woodruff Ave, Columbus, OH 43210}
\affiliation{Department of Astronomy, The Ohio State University, 140 W 18th Ave, Columbus, OH 43210}
 


\begin{abstract}

I find that Type Ia supernovae (\sneia) with bimodal nebular emission profiles occur almost exclusively in massive ($\Mstar \gtrsim 10^{11}~\Msun$) galaxies with low star-formation rates (SFR~$\lesssim 0.5~\Msun$/yr). The bimodal profiles are likely produced by two white dwarfs that exploded during a merger or collision, supported by a correlation between the peak-to-peak velocity separation (\vsep) and the \snia peak luminosity (\Mv) which arises naturally from more massive white dwarf binaries synthesizing more \fsNi during the explosion. The distributions of \sneia with and without bimodal nebular lines differ in host mass, SFR, and specific SFR with K-S test probabilities of $3.1\%$, $0.03\%$, and $0.02\%$, respectively. Viewing angle effects can fully explain the \sneia in quiescent hosts without bimodal emission profiles and the dearth of merger/collision driven \sneia in star-forming hosts requires at least two distinct progenitor channels for normal \sneia. $30-40\%$ of all \sneia originate from mergers or collisions depending on how cleanly host environment distinguishes progenitor scenarios. Existing models for white dwarf mergers and collisions broadly reproduce the \vsep--\Mv correlation and future analyses may be able to infer the masses/mass-ratios of merging white dwarfs in external galaxies. 

\end{abstract}



\section{Introduction} \label{sec:intro}

Type Ia supernovae (\sneia) are the thermonuclear explosion of white dwarf (WD) stars \citep{hoyle1960}. Since WDs do not spontaneously ignite, donation of matter from a companion is needed to destabilize the WD. The nature of the companion has been the source of extensive debate (see \citealp{maoz2014} and  \citealp{jha2019} for reviews). \sneia also produce the majority of iron-group elements in the Universe \citep{iwamoto1999, kobayashi2020} and the exact nucleosynthetic yields depend sensitively on the density and chemical composition of the core \citep[e.g., ][]{seitenzahl2013, boos2021, gronow2021}, which in turns depends on how the WD explodes as a \snia.

Progenitors are generally divided into 2 categories depending on the nature of the companion. The single-degenerate (SD) scenario invokes a non-degenerate companion such as a main-sequence or red giant star \citep{whelan1973, nomoto1982}, whereas the double-degenerate (DD) scenario has a second WD as the companion star \citep[e.g., ][]{tutukov1979, iben1984}. The SD scenario predicts several observational signatures from the SN ejecta interacting with the companion \citep[e.g., ][]{wheeler1975, marietta2000, kasen2010, boehner2017} or its wind \citep[e.g., ][]{panagia2006, margutti2012}. Searches for these signatures consistently produce null results (e.g., \citealp{mattila2005, leonard2007, schaefer2012, shappee2013a, chomiuk2016, maguire2016, dubay2022, tucker2024}, see \citealp{liu-rev-2023} for a review). \citet{tucker2020}, for example, limit the fraction of SD \sneia to $\lesssim 10\%$.

The non-detections of SD companions implies that double-WD binaries produce most \sneia. The question becomes \textit{how} the two WDs evolve to a detonation. Gravitational-wave emission and tidal forces can slowly reduce the orbital separation until mass-transfer or disruption occurs \citep[e.g., ][]{carvalho2022}. Mergers can explode during coalescence \citep[`violent' mergers; e.g., ][]{guillochon2010,pakmor2010, pakmor2012, pakmor2013} or after some time delay spanning $\sim 10^2-10^6$~years \citep[e.g., ][]{ji2013, becerra2018, neopane2022}. Dynamics in triple- and quadruple-star systems can induce mergers or even direct collisions \citep{thompson2011, katz2012, kushnir2013, shappee2013, pejcha2013, antognini2014, fang2018, hamers2022} but these scenarios struggle to reproduce the total \snia rate \citep[e.g., ][]{toonen2018, hamers2018}. Double detonations in double-WD binaries \citep[e.g., ][]{livne1990, shen2014} show promise for reproducing normal \sneia \citep[e.g., ][]{shen2018, polin2019}, but current models over-predict the diversity of \snia light curves \citep[e.g., ][]{shen2021, collins2022}. 

Host-galaxy and environmental analyses, which indirectly trace \snia progenitors, consistently find a diversity of explosion sites \citep[e.g., ][]{hakobyan2020, cronin2021}. There are correlations between the local environment and \sn properties \citep{childress2013, pan2015, meng2023} but the underlying origin is unknown. Accounting for these effects in cosmological analyses remains empirical (e.g., the `mass-step' correction, \citealp{kelly2010, sullivan2010}) and recent analyses suggest distinct progenitors in young and old stellar populations \citep[e.g., ][]{wang2013, rigault2020} with a $\sim 0.1-0.2$~mag intrinsic luminosity difference \citep{briday2022}. 

Yet connecting these environmental correlations to intrinsic ignition conditions or progenitor systems has proven difficult, primarily due to the explosion physics inherent to \sneia. Exploding a $\sim 1~\Msun$ star comprised of carbon and oxygen naturally produces a bright transient and layered ejecta with iron-group elements in the core surrounded by intermediate-mass elements from incomplete Si-burning \citep{seitenzahl2017}. Sky surveys and follow-up programs have obtained light curves and spectra for hundreds of \sneia in the weeks around maximum light \citep[e.g., ][]{blondin2012, silverman2012, morrell2024} but these observations only probe the outer ejecta \citep[e.g., ][]{mazzali2014}.

Nebular-phase spectra, obtained $\gtrsim 200$~days after explosion when the ejecta becomes optically thin, provide a direct view to the central regions of the explosion. These spectra are dominated by high-velocity forbidden emission lines from iron-group elements (Co, Ni, Fe; e.g., \citealp{wilk2020, kwok2023}) heated by radioactive-decay energy \citep{seitenzahl2009}. Most nebular spectra show smooth emission-line profiles \citep[e.g., ][]{maguire2018} with correlations between early-time and late-time ejecta velocities suggesting minor ejecta asymmetries \citep[e.g., ][]{maeda2010, liu2023}. Yet \citet{dong2015} discovered a subset of \sneia that show significantly asymmetric, even bimodal or double-peaked, nebular emission lines. \citet{vallely2020} later used an expanded sample to show that the \sneia with bimodal profiles have below-average luminosities but otherwise resemble spectroscopically-normal \sneia. 

In this Letter I connect these bimodal \sneia to exclusively old stellar populations. Moreover, the peak-to-peak velocity separation of the nebular profiles correlates with peak luminosity, a key prediction for WD mergers or collisions because total binary mass \Mtot corresponds to faster orbital velocities and more synthesized \fsNi. I outline the sample of \sneia searched for bimodality in \S\ref{sec:data}. Bimodal nebular profiles are connected to host-galaxy mass and star-formation rate (SFR) in \S\ref{sec:hosts} and I discuss these findings in the context of \sneia and their progenitors in \S\ref{sec:discuss}.

\begin{figure*}
    \centering
    \includegraphics[width=\linewidth]{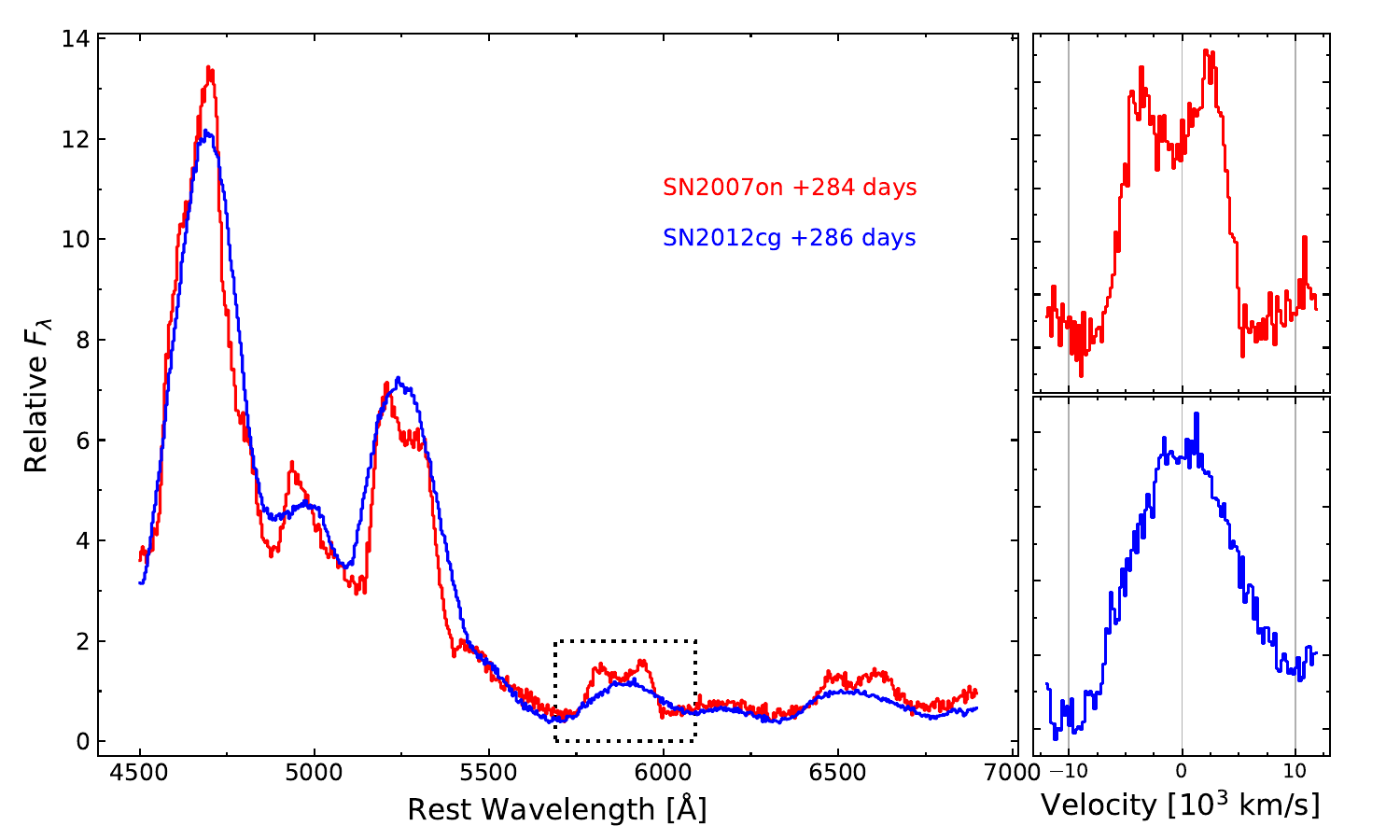}
    \caption{Example nebular spectra for \sneia with (SN~2007on, red) and without (SN~2012cg, blue) bimodal profiles. The dotted box marks the isolated \fCo{3}$\lambda5891$ feature shown in velocity space in the right panels.}
    \label{fig:spectra}
\end{figure*}

\begin{figure*}
    \centering
    \includegraphics[width=\linewidth]{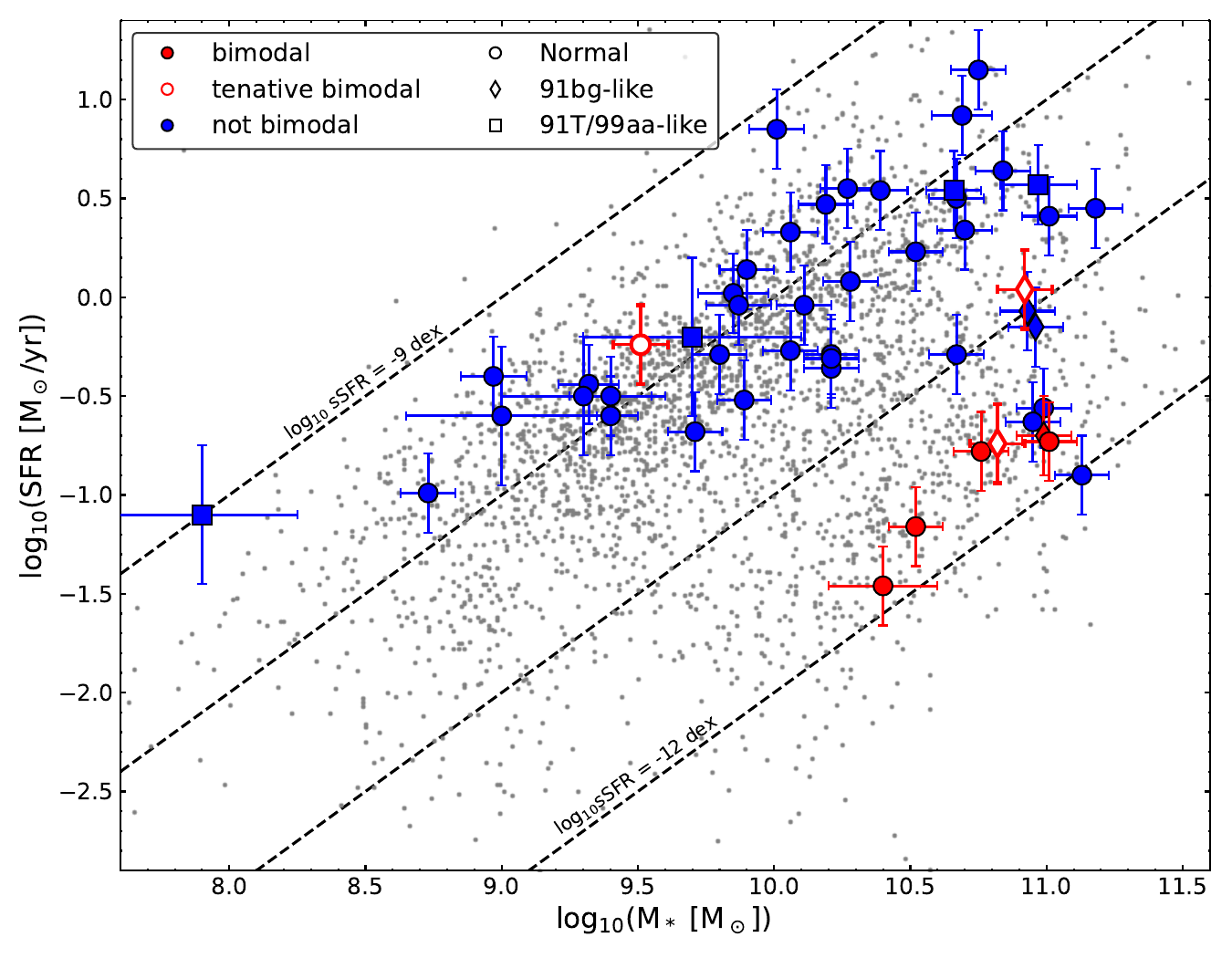}
    \caption{The host-galaxy masses and SFRs of bimodal \sneia (red) are compared to \sneia without bimodal nebular lines (blue). One host, NGC~1404, had 2 bimodal \sneia (2007on and 2011iv). Small grey points show a subset of galaxies from the z0MGS catalog \citep{leroy2019} for reference. The bimodal \sneia strongly prefer massive quiescent hosts.}
    \label{fig:mstar-sfmr}
\end{figure*}

\section{The Sample}\label{sec:data}

\begin{figure*}
    \centering
    \includegraphics[width=\linewidth]{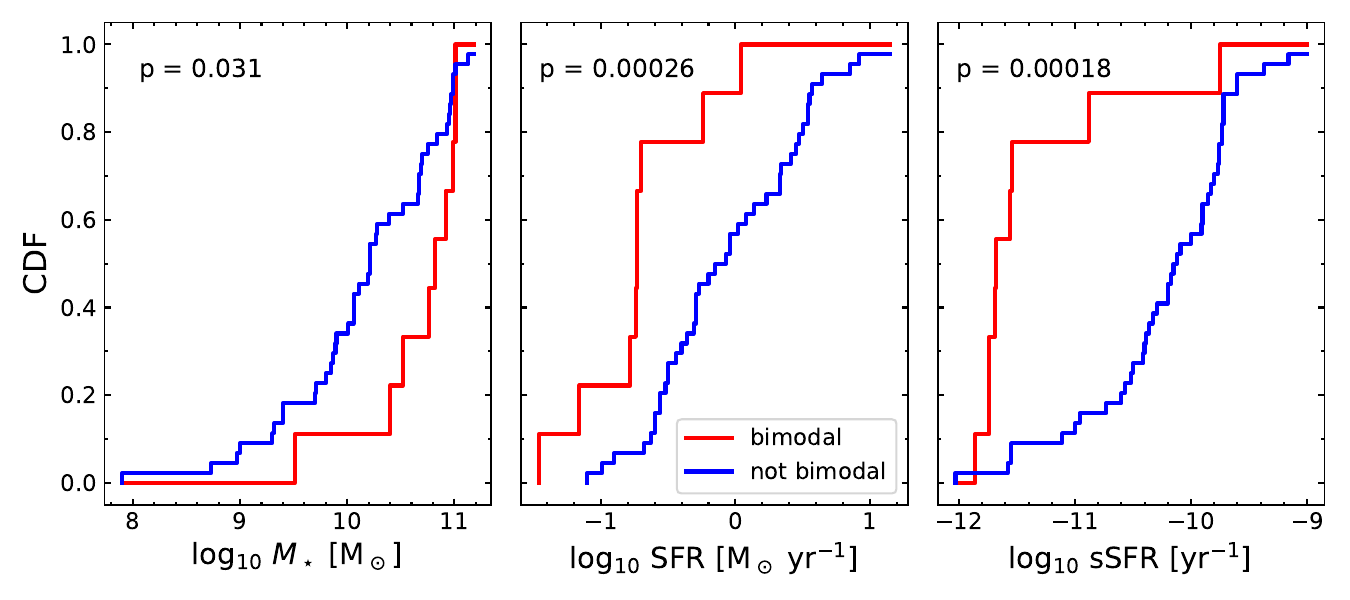}
    \caption{Cumulative distribution functions (CDFs) for \sneia with (red) and without (blue) bimodal nebular profiles. I show \Mstar (left), SFR (middle), and sSFR (right) including the $p$-value from the K-S test. The null hypotheses (the distributions are drawn from the same parent populations) can be rejected at the 99.9\% level for SFR and sSFR.}
    \label{fig:CDFs}
\end{figure*}

\citet{dong2015} and \citet{vallely2020} conducted dedicated searches for \sneia with bimodal nebular profiles which I use as an initial sample. Those works searched for bimodality in 48 \sneia using new and archival nebular spectra. The \sneia are classified as `bimodal' if the two velocity components are separated by more than their combined width (cf. Table~4 in \citealp{vallely2020}). Table~\ref{tab:allsneia} provides the basic SN and host-galaxy properties for the sample.

There are 6 \sneia in this sample that show bimodality in all three spectral features considered: the \fFe{2} blend at $\approx 5300~\AAA$, the isolated \fCo{3}$\lambda 5891$ line, and the \fFe{2}+\fCo{3} blend at $\approx 6600$~\AAA. Fig.~\ref{fig:spectra} shows the nebular spectrum of SN~2007on \citep{gall2018, mazzali2018} with bimodality in all three features. Included for comparison is the nebular spectrum of SN~2012cg \citep{silverman2012b, shappee2018, chakrahari2019} at a similar phase which is classified as  `not bimodal'. The narrow host-galaxy \ion{Na}{1} absorption feature has been subtracted from the spectrum of SN~2012cg and the spectra are aligned using the velocity shifts derived by \citet{vallely2020} to ease visual comparisons.

There are 3 additional \sneia for which I retain the `tentative bimodal' classification from \citet{vallely2020}. Two of these \sneia (2002er and 2016iuh) only show bimodal signatures in the \fCo{3}$\lambda5891$ feature. The spectra for these \sneia were obtained earlier in the nebular phase ($+214$ and $+164$~days, respectively) compared to the \sneia with bimodality detected in all three features (phases $\sim 250-350$~days). $^{56}$Co decays to $^{56}$Fe with a half-life of $\approx 77$~days so these differences may simply trace the conversion of Co into Fe. 

The third `tentative bimodal' \snia, 1986G, is more likely to be a contamination artifact than SNe~2002er and 2016iuh. This explosion suffered from significant host-galaxy extinction and the host \ion{Na}{1} absorption doublet falls atop the \fCo{3}$\lambda5891$ feature. The redder \fFe{2}+\fCo{3} feature also shows a strong absorption feature that might be due to over-subtraction of host-galaxy \Ha emission. Yet SN~1986G has similar properties to the other bimodal \sneia so I retain the `tentative bimodal' classification. All three of these \sneia have open markers in the corresponding figures.

Finally, I add the 5 \sneia with nebular spectra from \citet{graham2022} who searched for bimodality but did not find any detections. The combined sample from these 3 sources consists of 53 \sneia, 6 of which are confidently bimodal, three are tentatively bimodal, and the remaining 44 \sneia show no evidence for bimodal line profiles. 

The full sample spans a wide range of near-peak photometric and spectroscopic properties (cf. Fig.~7 in \citealp{vallely2020}). Most of them are spectroscopically normal but there are 5 underluminous 91bg-like \citep{filippenko1992b, leibundgut1993} and 5 overluminous 91T/99aa-like \citep{filippenko1992, phillips1992, garavini2004}, denoted with specific symbols in the corresponding figures. The 02cx-like/Iax \sne \citep{li2003, foley2013} are excluded as they likely originate from a different progenitor class and appear to never enter a true nebular phase \citep[e.g., ][]{foley2016, camacho-neves2023}. 

\section{Host Galaxies}\label{sec:hosts}

I compare the \sneia with and without bimodality to the stellar masses (\Mstar), star-formation rates (SFR), and specific SFR (${\rm{sSFR}\equiv\rm{SFR}/\Mstar}$) of their hosts. Most hosts have \Mstar and SFR measurements in the `$z=0$ Multi-wavelength Galaxy Synthesis' (z0MGS; \citealp{leroy2019}) catalog anchored to $\sim 700,000$ galaxies in the GALEX-SDSS-WISE Legacy Catalog \citep[GSWLC; ][]{salim2016}. Hosts for 4 of the \sneia in my sample are not included in \textsc{z0MGS} so we apply the same calibrations from \citet{leroy2019} to the GALEX and WISE photometry for these hosts to ensure consistency. Simply excluding these 4 \sneia, all of which are classified as `not bimodal' and explode in star-forming hosts, does not affect the conclusions.

Fig.~\ref{fig:mstar-sfmr} shows the galaxy masses and SFRs for the \snia hosts color-coded by bimodality, The bimodal \sneia occur almost exclusively in massive ($\logMstar \simeq 11$~dex), quiescent ($\logSFR \lesssim -0.5$~dex) galaxies. The exception is the tentatively bimodal SN~2002er which occurs in a star-forming host. Two of the bimodal \sneia, SNe~2007on and 2011iv, share the same host galaxy NGC~1404.

I use the Kolmogorov-Smirnov (K-S) test to quantify whether the host properties of bimodal and non-bimodal \sneia differ. Fig.~\ref{fig:CDFs} shows the cumulative distribution functions (CDFs) for \logMstar, \logSFR, and \logsSFR including the corresponding K-S $p$-values. The SFR and sSFR distributions in particular show that the null hypotheses of no correlation between host-galaxy property and bimodality can be rejected at ${>99.9\%}$ confidence. \citet{leroy2019} note that their calibrations for SFR (and thus sSFR) likely overestimate the true values for passive galaxies ($\logsSFR \lesssim -11$~dex) because star-forming galaxies dominate their calibration sample. Reducing the SFR for these galaxies would further separate the bimodal \sneia from the majority of \snia hosts.

The presence of \sneia without bimodal profiles in these massive ellipticals agrees with viewing-angle effects obscuring asymmetry in some systems. If I just consider \sneia with hosts with $\logsSFR \leq-10.8$~dex \citep{rigault2020}, $P_{\rm bim} =8/15~(53\%)$ of these \sneia show bimodal profiles. 
The observed fraction of bimodal \sneia $P_{\rm bim}$ is related to the critical viewing angle by

\begin{equation}\label{eq:theta}
    \theta_c = 90^\circ - \cos^{-1}(P_{\rm bim}/f),
\end{equation}

\noindent where $f$ is the `true' fraction of bimodal \sneia. Explosions from violent mergers and collisions are expected to exhibit strong asymmetries, especially in the inner ejecta probed by the nebular Fe and Co features \citep[e.g., ][]{pakmor2013, dong2015}. For $f\equiv1$, $\theta_c \simeq35^\circ$, and this requires that most, if not all, \sneia in $\logsSFR \lesssim -11$~dex galaxies are intrinsically bimodal.

The dearth of bimodal events in star-forming hosts is intriguing because these galaxies have a mixture of young and old stellar populations. SN~2002er is the sole (tentative) bimodal \snia in a star-forming host in Fig.~\ref{fig:mstar-sfmr} and viewing-angle effects likely obscures bimodality in an additional $1-2$ events (cf. Eq.~\ref{eq:theta}). This implies ${(2\pm1)/38\sim 5\%}$ (95-percentile confidence interval ${1-16\%}$) of \sneia in star-forming hosts are produced by mergers, using the same sSFR cutoff from \citet{rigault2020}.

The separation by host sSFR allows a relatively clean estimate of the true fraction of all \sneia originating from merging WD binaries. Of the \sneia in my sample, 9/53 ($\simeq 17\%$) show evidence for bimodality and accounting for the $\sim$half that will not be classified as bimodal due to viewing-angle effects raises this fraction to $\sim 34\%$. \sneia in quiescent galaxies, including bimodal \sneia, are systematically fainter \citep[e.g., ][]{kelly2010, sullivan2010, gupta2011, rigault2013} so there is likely some bias against late-time spectra for these events. For comparison, $\sim 35-40\%$ of the \snia hosts analyzed by \citet{rigault2020} have $\logsSFR \lesssim -11$~dex although stellar population does not perfectly separate the two populations \citep{briday2022}. With these considerations in mind, the intrinsic fraction of bimodal \sneia relative to all \sneia is probably 30--40\%. 

Finally, I do not find an unambiguous connection to the spectroscopic 91bg-like subclass of \sneia (diamonds in Fig.~\ref{fig:mstar-sfmr}), which also have below-average peak luminosities \citep[e.g., ][]{filippenko1992b, taubenberger2008, hachinger2009} and a strong preference for passive hosts \citep[e.g., ][]{barkhudaryan2019}. If 91bg-like classification is inclination dependent similar to bimodality, we would expect consistency between the bimodal and 91bg-like classifications. Common symmetry axes would produce correlated outcomes (bimodal$\Leftrightarrow$91bg-like) whereas orthogonal axes would produce an anti-correlation (bimodal$\neq$91bg-like). Neither of these are observed, as Fig.~\ref{fig:mstar-sfmr} shows all 4 possible combinations of these categories. Either there are different axes of symmetry between the inner and outer ejecta or different physical processes divide these empirical categories \citep[e.g., ][]{dhawan2017}.

\section{Implications for SN Ia Progenitors}\label{sec:discuss}

\begin{figure*}
    \centering
    \includegraphics[width=\linewidth]{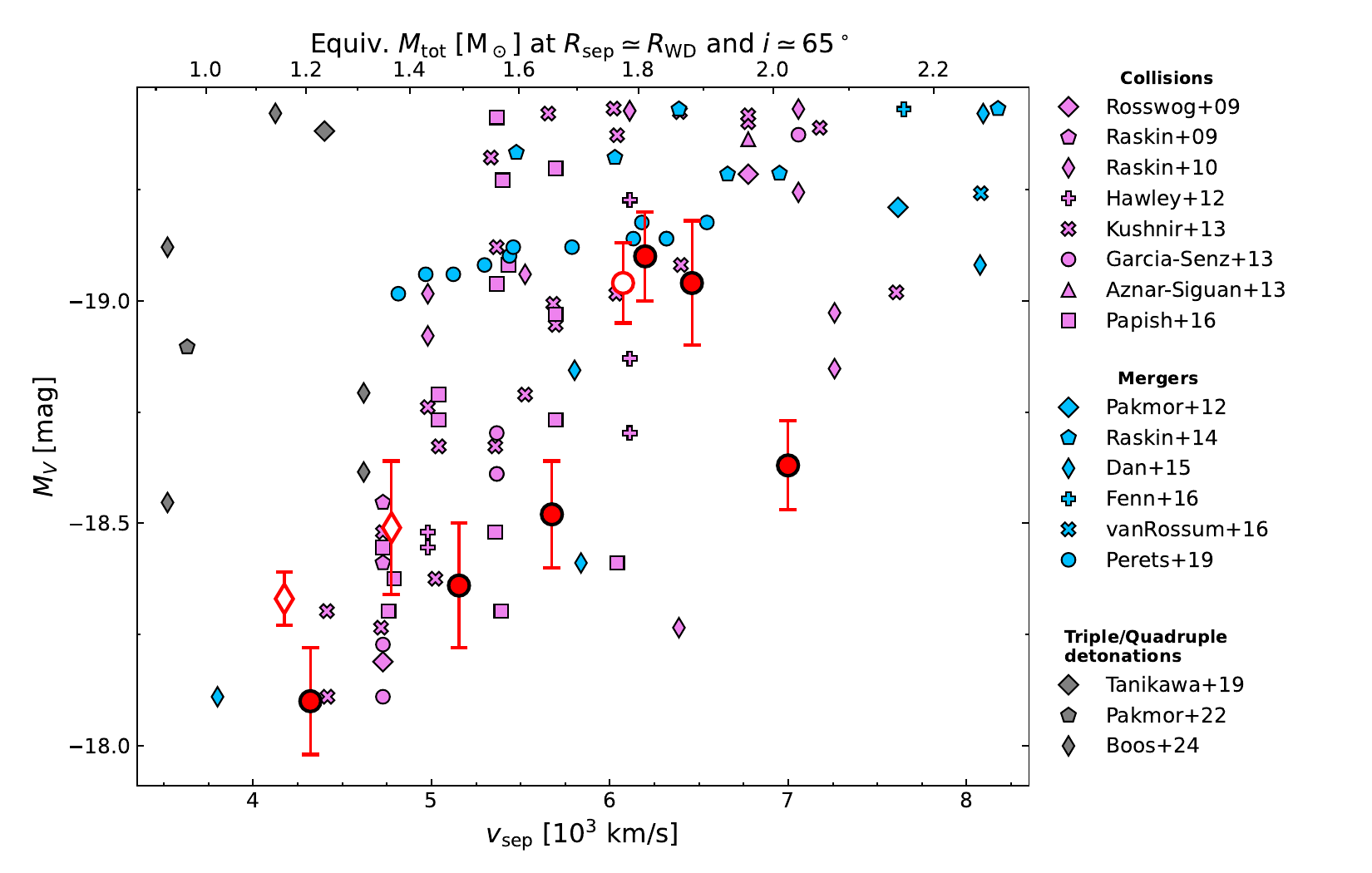}
    \caption{Correlation between the bimodal peak-to-peak velocity separation \vsep and peak $V$-band magnitude. Observations have the same markers and colors as in Fig.~\ref{fig:mstar-sfmr}. The top axis shows the total binary mass if $v_{\rm orb} \sim 1.1\times \vsep$ corresponding to an average inclination of $i\simeq65^\circ$. I use the synthesized \Mni to estimate \Mv for models where two WDs explode, including collisions (purple, \citealp{rosswog2009, raskin2009, raskin2010, hawley2012, kushnir2013, garciasenz2013, aznarsiguan2013, papish2016}), mergers (blue; \citealp{pakmor2012, raskin2014, dan2015, fenn2016, vanRossum2016, perets2019}), and triple- or quadruple-detonations (gray, \citealp{tanikawa2019, pakmor2022, boos2024}). See \S\ref{sec:discuss} for details.}
    \label{fig:vsep-corr}
\end{figure*}

The dependence on sSFR explains correlations between \sneia and their hosts discovered over the past decade including the `step-correction' parameterized by host mass, SFR, or sSFR \citep[e.g., ][]{kelly2010, sullivan2010, lampeitl2010, smith2012, wang2013, pan2014, roman2018, rigault2020, kang2020, kelsey2021, briday2022, wiseman2022, wiseman2023, larison2024, ginolin2024}. \citet{vallely2020} find that bimodal \sneia are $\delta m\sim 0.3$~mag fainter than the population average, slightly higher than the $\delta m\sim 0.2$~mag offset inferred by \citet{briday2022} from a large sample of \sneia and host-galaxy properties. Yet the bimodal \sneia already have at least one axis of asymmetry so viewing-angle effects may account for the slightly larger $\delta m$ calculated by \citet{vallely2020} if luminosity and inclination are correlated.

The peak-to-peak velocity separations \vsep in bimodal \sneia exceed pre-explosion turbulent motions in the WD ($\sim 100~\kms$; \citealp{hoeflich2002, kim2013}) and the preference for roughly equal-height peaks in the bimodal profiles also disfavors internal explosive asymmetries such as deflagration plumes \citep[e.g., ][]{seitenzahl2013} or off-center detonations \citep[e.g., ][]{chamulak2012}. Instead, multi-dimensional simulations of violent mergers \citep{pakmor2013} and collisions \citep{rosswog2009} show the highly-asymmetric \fsNi distributions required by the bimodal line profiles.

The correlation between \vsep and the peak $V$-band magnitude \Mv shown in Fig.~\ref{fig:vsep-corr} further suggests a merger origin, where \vsep measures the separation between the two velocity components in the nebular spectra and \Mv is a reliable proxy for the bolometric luminosity \citep[e.g., ][]{arnett1982}. The Spearman correlation coefficient for \vsep and \Mv in Fig.~\ref{fig:vsep-corr} is $r=-0.75$ with a statistically significant $p$-value of $p=0.020$. The negative correlation coefficient corresponds to a \emph{positive} correlation between \vsep and peak luminosity because the photometric magnitude system is inverted. This correlation is naturally produced if more massive WD binaries with faster orbits synthesize more \fsNi during the explosion, producing more radioactive decay power and brighter light curves. 

The old stellar populations, bimodal nebular profiles, and \vsep--\Mv correlation are all consistent with the explosion of merging double-WD binaries. The top axis of Fig.~\ref{fig:vsep-corr} converts \vsep into an estimate for the total binary mass \Mtot assuming equal-mass WDs. The inclination angle $i$ is unknown so I assume $\vorb \simeq 1.1\times \vsep$ corresponding to $i \approx 90-\theta_c \simeq65^\circ$. There is likely a slight observational bias towards identifying edge-on/high inclination systems. Equal(-ish) WD masses are supported by the bimodal \fCo{3} emission lines having similar fluxes, where the  \fCo{3} line flux $F_{\rm [Co~III]}\propto \Mni$ \citep{childress2015} and $\Mni\propto$ WD mass \citep[e.g., ][]{shen2018}.

I use the WD cooling tracks of \citet{bedard2020} to estimate WD radii (\Rwd) from their masses assuming thick H layers and a cooling age of 5~Gyr, although adopting anywhere between  $1-10$~Gyr produces similar results. To estimate \Mv from \Mni (a readily-provided quantity from most simulations), I combine the relation between the light curve `stretch' parameter \sbv \citep{burns2011} and \Mni from \citet{scalzo2019} with the \sbv-\Mv calibration derived by \citet{burns2018}. This allows for direct comparison between models and observations even for simulations lacking radiative transfer computations. 

Overall the collision and merger models agree well with the data in Fig.~\ref{fig:vsep-corr} assuming the separation at explosion is $R_{\rm sep} \simeq (R_1 + R_2)/2$ (i.e., partially merged). This agrees with the bimodal Co and Fe emission lines overlapping in velocity (cf. Fig.~\ref{fig:spectra}) yet retaining distinct cores \citep[e.g., ][]{pakmor2013, mazzali2018}. Relating \vsep to the pre-explosion orbital velocities assumes the binary is immediately unbound during explosion, but in reality these velocities are lower limits because self-gravitation from the ejecta reduces \vsep by $\sim 10-20\%$ \citep[e.g., ][]{braudo2024}, depending on the explosion energy and ejecta mass. Adopting different orbital separations at explosion moves the models horizontally in Fig.~\ref{fig:vsep-corr} and different viewing angles will likely shift the models along both axes because simulations suggest inclination-dependent light curves and spectra \citep[e.g., ][]{rosswog2009, pakmor2012}. Mergers must ignite during or soon after coalescence to produce the \fsNi asymmetries, and any merger scenario must also reproduce the non-explosive merger remnants \citep[e.g., ][]{cheng2020, kilic2021, wu2022}. 

Collisions depend on the impact parameter $b$ where off-axis or grazing collisions reduce \Mni or potentially prevent ignition altogether \citep[e.g., ][]{glanz2023}. Direct collisions are too rare to reproduce the \snia rate by a factor of $\gtrsim 100$ \citep[e.g., ][]{hamers2013, toonen2018}, so they are unlikely to produce a significant fraction of the bimodal \sneia. Triple systems more broadly may play an important role in driving WD mergers \citep[e.g., ][]{hamers2022}.

The triple- and quadruple-detonation models \citep{papish2015} use a double-detonation on the primary to compress and ignite the secondary (triple detonations) or ignite a second double-detonation (quadruple detonations). These models have too low \vsep to match the data in Fig.~\ref{fig:vsep-corr} because they explode at larger separations and thus lower orbital velocities. However, this may depend on how the model binary is driven to explosion (\citealp{burmester2023}), when the He accretion layer ignites,  or if a kick is imparted to the primary's ejecta by the detonation of the secondary. \\

\vspace{0.5cm}

This rare electromagnetic observable to merging compact objects constrains key aspects of binary population synthesis models \citep[e.g., ][]{yungelson2017, shen2017},  delay-time distributions \citep[e.g., ][]{maoz2010, meng2012}, and the stochastic gravitational-wave background \citep[e.g., ][]{yu2020, zou2020, yoshida2021}.
Many questions remain about how exactly the double WD binary is driven to explosion but these results confirm that merging double-WD binaries can and do produce spectroscopically-normal \sneia around maximum light. Moreover, this requires at least 2 separate progenitor channels for producing \sneia. The connection to host-galaxy sSFR allows for early identification of likely bimodal \sneia for detailed study. 
\vspace{1cm}

\software{
astropy \citep{astropy}; numpy \citep{numpy1, numpy2}; matplotlib \citep{matplotlib}; lmfit \citep{lmfit}; scipy \citep{scipy}; spectres \citep{spectres}; emcee \citep{emcee}; pandas \citep{pandas}}
\\

\section*{Data Availability}
All data are taken from the literature. Host properties for the \snia sample are included in Table~\ref{tab:allsneia} for reference.

\section*{Acknowledgments}

Many thanks to Adam Leroy, Chris Kochanek, Ben Shappee, Kris Stanek, Todd Thompson, Ken Shen, and Noam Soker for useful discussions. I thank the referee for their constructive suggestions.

\begin{table*}
    \centering
    \footnotesize
\begin{tabular}{llllrrrr}
\toprule
        Name & Bimodal? & Subtype &                       Host &   \logMstar [dex] &  \logSFR [dex] \\
\midrule
       1981B &        N &    norm &                    NGC~4536 &  $10.19\pm0.10$ &  $  0.47 \pm 0.20$ \\
       1986G &        T &    91bg &                    NGC~5128 &  $10.92\pm0.10$ &  $  0.04 \pm 0.20$ \\
       1989B &        N &    norm &                        M~66 &  $10.67\pm0.10$ &  $  0.50 \pm 0.20$ \\
       1990N &        N &    norm &                    NGC~4639 &  $10.21\pm0.10$ &  $ -0.29 \pm 0.20$ \\
       1991T &        N &     91T &                    NGC~4527 &  $10.66\pm0.10$ &  $  0.54 \pm 0.20$ \\
      1998aq &        N &    norm &                    NGC~3982 &  $ 9.90\pm0.10$ &  $  0.14 \pm 0.20$ \\
      1998bu &        N &    norm &                        M~96 &  $10.67\pm0.10$ &  $ -0.29 \pm 0.20$ \\
      1999aa &        N &     91T &                    NGC~2595 &  $10.97\pm0.14$ &  $  0.57 \pm 0.20$ \\
      1999by &        N &    91bg &                    NGC~2841 &  $10.93\pm0.10$ &  $ -0.07 \pm 0.20$ \\
      2000cx &        N &    norm &                    NGC~0524 &  $10.99\pm0.10$ &  $ -0.56 \pm 0.20$ \\
      2002dj &        N &    norm &                    NGC~5018 &  $10.95\pm0.10$ &  $ -0.63 \pm 0.20$ \\
      2002er &        T &    norm &                   UGC~10743 &  $ 9.51\pm0.10$ &  $ -0.24 \pm 0.20$ \\
      2003du &        N &    norm &                    UGC~9391 &  $ 8.73\pm0.10$ &  $ -0.99 \pm 0.20$ \\
      2003hv &        Y &    norm &                    NGC~1201 &  $10.40\pm0.20$ &  $ -1.46 \pm 0.20$ \\
      2003gs &        Y &    91bg &                    NGC~0936 &  $10.99\pm0.10$ &  $ -0.70 \pm 0.20$ \\
      2004bv &        N &    norm &                    NGC~6907 &  $10.69\pm0.11$ &  $  0.92 \pm 0.20$ \\
      2004eo &        N &    norm &                    NGC~6928 &  $11.18\pm0.10$ &  $  0.45 \pm 0.20$ \\
      2005am &        Y &    norm &                    NGC~2811 &  $10.76\pm0.10$ &  $ -0.78 \pm 0.20$ \\
      2005cf &        N &    norm &              MCG~-01-39-003 &  $ 9.40\pm0.15$ &  $ -0.50 \pm 0.20$ \\
      2007af &        N &    norm &                    NGC~5584 &  $ 9.85\pm0.13$ &  $  0.02 \pm 0.20$ \\
      2007le &        N &    norm &                    NGC~7721 &  $10.11\pm0.10$ &  $ -0.04 \pm 0.20$ \\
      2007on &        Y &    norm &                    NGC~1404 &  $11.01\pm0.10$ &  $ -0.73 \pm 0.20$ \\
       2008Q &        N &    norm &                    NGC~0524 &  $10.99\pm0.10$ &  $ -0.56 \pm 0.20$ \\
      2011by &        N &    norm &                    NGC~3972 &  $ 9.71\pm0.10$ &  $ -0.68 \pm 0.20$ \\
      2011fe &        N &    norm &                       M~101 &  $10.39\pm0.10$ &  $  0.54 \pm 0.20$ \\
      2011iv &        Y &    norm &                    NGC~1404 &  $11.01\pm0.10$ &  $ -0.73 \pm 0.20$ \\
      2012cg &        N &    norm &                    NGC~4424 &  $ 9.89\pm0.10$ &  $ -0.52 \pm 0.20$ \\
      2012fr &        N &    norm &                    NGC~1365 &  $10.75\pm0.10$ &  $  1.15 \pm 0.20$ \\
      2012hr &        N &    norm &                ESO~121-G026 &  $10.52\pm0.10$ &  $  0.23 \pm 0.20$ \\
      2013aa &        N &    norm &                    NGC~5643 &  $10.06\pm0.10$ &  $  0.33 \pm 0.20$ \\
      2013dy &        N &    norm &                    NGC~7250 &  $ 8.97\pm0.12$ &  $ -0.40 \pm 0.20$ \\
      2013gy &        N &    norm &                    NGC~1418 &  $10.21\pm0.10$ &  $ -0.36 \pm 0.20$ \\
       2014J &        N &    norm &                        M~82 &  $10.01\pm0.10$ &  $  0.85 \pm 0.20$ \\
      2014bv &        Y &    norm &                    NGC~4386 &  $10.52\pm0.10$ &  $ -1.16 \pm 0.20$ \\
       2015F &        N &    norm &                    NGC~2442 &  $10.84\pm0.10$ &  $  0.64 \pm 0.20$ \\
       2015I &        N &    norm &                    NGC~2357 &  $10.06\pm0.10$ &  $ -0.27 \pm 0.20$ \\
     2016coj &        N &    norm &                    NGC~4125 &  $11.13\pm0.10$ &  $ -0.90 \pm 0.20$ \\
     2016fnr &        N &    norm &                   UGC~10502 &  $10.70\pm0.10$ &  $  0.34 \pm 0.20$ \\
     2016iuh &        T &    91bg &                   UGC~07367 &  $10.82\pm0.10$ &  $ -0.74 \pm 0.20$ \\
     2016gxp &        N &     91T &                    NGC~0051 &  $ 9.70\pm0.40$ &  $ -0.20 \pm 0.40$ \\
     2017hjw &        N &    norm &                   UGC~03245 &  $10.28\pm0.10$ &  $  0.08 \pm 0.20$ \\
     2017cbv &        N &     91T &                    NGC~5643 &  $10.06\pm0.10$ &  $  0.33 \pm 0.20$ \\
     2017ckq &        N &    norm &                ESO~437-G056 &  $10.21\pm0.10$ &  $ -0.31 \pm 0.20$ \\
     2017erp &        N &    norm &                    NGC~5861 &  $10.27\pm0.10$ &  $  0.55 \pm 0.20$ \\
     2017fzw &        N &    91bg &                    NGC~2217 &  $10.96\pm0.10$ &  $ -0.15 \pm 0.20$ \\
      2018gv &        N &    norm &                    NGC~2525 &  $ 9.87\pm0.12$ &  $ -0.04 \pm 0.20$ \\
      2018oh &        N &    norm &                   UGC~04780 &  $ 9.32\pm0.11$ &  $ -0.44 \pm 0.20$ \\
      2018yu &        N &    norm &                    NGC~1888 &  $11.01\pm0.10$ &  $  0.41 \pm 0.20$ \\
 ASASSN-14jg &        N &    norm &  J2333--6034$^a$ &   $9.80 \pm 0.10$ &   $-0.29 \pm 0.20$ \\
 ASASSN-15uh &        N &     91T &                KUG~0925+693 &   $7.90 \pm 0.35$ &  $ -1.10 \pm 0.35$ \\
 ASASSN-16lx &        N &    norm &                     IC~0607 &   $9.00 \pm 0.35$ &   $-0.60 \pm 0.35$ \\
 ASASSN-17cz &        N &    norm &  J1750--0148$^b$ &   $9.30 \pm 0.30$ & $ -0.50 \pm 0.30$ \\
 ASASSN-17pg &        N &    norm &                    NGC~3901 &   $9.40 \pm 0.10$ &  $ -0.60 \pm 0.20$ \\
\bottomrule
\end{tabular}
    \caption{SN and host properties for the sample. Entries in the `Bimodal?' column are one of: (Y)es, (N)o, or (T)entative. $^q$WISEA J233312.25--603420.6. $^b$WISEA J175030.58--014802.8.}
    \label{tab:allsneia}
\end{table*}

%




\bibliography{ref}{}
\bibliographystyle{aasjournal}



\end{document}